\documentclass{tufte-handout}

\usepackage{enumitem}

\usepackage{amsmath}
\usepackage{pifont}
\usepackage{url}
\makeatletter
\def\url@leostyle{%
  \@ifundefined{selectfont}{\def\UrlFont{\sf}}{\def\UrlFont{\footnotesize\ttfamily}}}
\makeatother
\usepackage{graphicx}
\setkeys{Gin}{width=\linewidth,totalheight=\textheight,keepaspectratio}
\graphicspath{{graphics/}}

\title{\Large{\bf One, None and One Hundred Thousand Profiles} \\\footnotesize{Re-imagining the Pirandellian Identity Dilemma in the Era of Online Social Networks}}
\author{Alberto Pepe, Spencer Wolff \& Karen Van Godtsenhoven\marginnote{A.P. is at Harvard University. S.W. is at Yale University. K.V.G. is at Ghent University}}
\date{ }  % if the \date{} command is left out, the current date will be used

\usepackage{booktabs}

\usepackage{units}

\usepackage{fancyvrb}
\fvset{fontsize=\normalsize}

\usepackage{multicol}

\usepackage{lipsum}

\begin{document}

\maketitle% this prints the handout title, author, and date

\section{Abstract}
\marginnote{Note: This is a preprint. An abridged version of this paper will be presented/performed under the title ``Identity dilemmas on Facebook'' at the  Symposium on the Dynamics of the Internet and Society ``A Decade in Internet Time'' to be held Wednesday 21 - Saturday 24 September 2011 at the Oxford Internet Institute, University of Oxford, UK.} \textit{Uno, Nessuno, Centomila} (``One, No One and One Hundred Thousand'') is a classic novel by Italian playwright Luigi Pirandello. Published in 1925, it recounts the tragedy of Vitangelo Moscarda, a man who struggles to reclaim a coherent and unitary identity for himself in the face of an inherently social and multi-faceted world. What would Moscarda identity tragedy look like today? In this article we transplant Moscarda's identity play from its offline setting to the contemporary arena of social media and online social networks. With reference to established theories on identity construction, performance, and self-presentation, we re-imagine how Moscarda would go about defending the integrity of his selfhood in the face of the discountenancing influences of the online world.

\vspace{10em}

\begin{quote}
\textit{You have one identity. [...] The days of you having a different image for your co-workers and for the other people you know are probably coming to an end pretty quickly. [...] Having two identities for yourself is an example of a lack of integrity.}
\flushright{Mark Zuckerberg}
\end{quote}

\vspace{2em}

\section{Introduction}

With the growing permeation of online social networks in our everyday life, scholars have become interested in the study of novel forms of identity construction, performance, spectatorship and self-presentation onto the networked medium. This body of research builds upon a rich theoretical tradition on identity constructivism, performance and (re)presentation of self. With this article we attempt to integrate the work of Italian playwright Luigi Pirandello into this tradition.

Pirandello's classic 1925 novel \textit{Uno, Nessuno, Centomila} (``One, No One and One Hundred Thousand") recounts the tragedy of a man who struggles to reclaim a coherent identity for himself in the face of an inherently social and multi-faceted world. Via an innocuous observation of his wife, the protagonist of the novel, Vitangelo Moscarda, discovers that his friends' perceptions of his character are not at all what he imagined and stand in glaring contrast to his private self-understanding.  In order to upset their assumptions, and to salvage some sort of stable identity, he embarks upon a series of carefully crafted social experiments. 

Though the novel's story transpires in a pre-digital age, the volatile play of identity that ultimately destabilizes Moscarda has only increased since the advent of online social networks. The constant flux of communication in the online world frustrates almost any effort at constructing and defending unitary identity projections. Popular social networking sites, such as Facebook and MySpace, offer freely accessible and often jarring forums in which widely heterogeneous aspects of one's life---that in Moscarda's era could have been scrupulously kept apart---precariously intermingle. Disturbances to our sense of a unified identity have become a matter of everyday life. 

Pirandello's prescient novel offered readers in its day the contours of an identity melee that would unfurl on the online arena some 80 years later. Reading Uno, Nessuno e Centomila in light of established and emerging theories on identity construction, we re-imagine how Moscarda would go about defending the integrity of his selfhood in the face of the discountenancing influences of the online world. In other words, we address the following question: How would Moscarda's tragedy play out in the inherently networked world of today? This article hopes to shed light on contemporary dilemmas of identity constructivism and self-representation while simultaneously re-evaluating one of the most celebrated works of one of Italy's profoundest thinkers on identity and personhood. 

This article begins with a synopsis of Pirandello's novel, presented in the next section. The remainder of the article analyses in detail four crucial episodes of the novel that we attempt to reframe from their original offline, place-based setting, to the contemporary online stage, in which offline, physical interactions necessarily intersect with the networked digital medium. 

\section{The drama of Vitangelo Moscarda}

\textit{Uno, Nessuno e Centomila} is Pirandello's last and possibly most complete novel. Published in 1925 in eight chapters (books), it is the result of nearly 20 years of intense, yet oft-interrupted, work. Despite this long and difficult period of preparation, the novel is a lucid compendium of the central themes in Pirandello's thought. In this section, we present a synopsis of the novel and its central themes, focusing on the four episodes that are further discussed in the following sections\footnote{All excerpts (and relative page numbers) of One, No One and One Hundred Thousand in this and following sections refer to the edition and translation of the book listed in the references (Pirandello, 1990).}. 

One, No one and One Hundred Thousand tells the tragedy of Vitangelo Moscarda, ``a man whose great aim was to see clearly and be truly himself" (p. xiii). In the narrative style of the \textit{bildungsroman}, Moscarda recounts his troubled experience of identity as a first-person narrator. The opening lines of the novel (p.3) set the stage for the entire novel:

\begin{quote}
\guillemotleft What are you doing?\guillemotright, my wife asked, seeing me linger, unusually in front of the mirror.\linebreak
\guillemotleft Nothing\guillemotright, I replied. \guillemotleft  Just looking at myself, at my nose, here, inside this nostril. When I press it, I feel a little pain\guillemotright. \linebreak
My wife smiled and said: \guillemotleft  I thought you were looking to see which way it tilts\guillemotright. \linebreak
I wheeled around like a dog whose tail has been stepped on. \linebreak
\guillemotleft  Tilts? My nose?\guillemotright
\end{quote}

With this sudden, unexpected, and accidental realization - that his nose tilts - Moscarda begins his dramatic journey to discover his own self. The tilted nose is the ``first germ of the sickness" (p. 7) through which Moscarda begins to question his own public image as he apprehends that his acquaintances have constructed, in their own imagination, a Moscarda persona that is likely to be different from his own: ``I believed everyone saw me as a Moscarda with a straight nose, whereas everyone saw a Moscarda with a bent nose" (p. 6) and ``I was obsessed by the thought that for others I was not what till now, privately, I had imagined myself to be" (p. 7). This realization soon becomes an identity dilemma - ``if for the others I was not the one I had always believed I was for myself, who was I?" (p. 12) - that culminates in a number of experiments with his own identity.   

As a first experiment, Moscarda tries to experience his own persona from the point of view of an outsider: ``I wanted to be alone in a quite unusual, new way, [...] namely, without myself and, in fact, with an outsider present" (p. 11). He sets himself before a mirror with the intent to watch himself ``live" from the outside; to catch himself in the most natural actions and expressions. This initial experiment is the first in a number of ``necessary follies" that Moscarda performs to explore the public perception of his own persona and identity. It is discussed more in detail and reframed in the arena of online social networks in the next section, ``The mirror experiment".

So-who is Moscarda? He is the director of a bank in the small town of Richieri-a bank he inherited from his father but that he knows very little about: ``I had two faithful friends, Sebastiano Quantorzo and Stefano Firbo, to handle my affairs after the death of my father, who, though he tried in every way, had never succeeded in making me accomplish anything; except taking a wife" (p. 4). Moscarda's first experiment helps him throw some light on his past and on the public's perception of his business persona. At work, he comes to realize, for example, that most people regard him (and his father) as a ``usurer", rather than a ``banker". In the family, he begins to understand the undertones of the nickname (``Geng\`{e}") given to him by his wife Dida. ``Ah, yes- for the others a usurer; [...] for my wife, Dida, a fool." (p. 57). A surprising mosaic of identities slowly unveils itself before Moscarda, and his aim becomes the deconstruction of this mosaic and every connotation associated with it: ``I decided to discover who I was at least for those closest to me, my so-called acquaintances, and to amuse myself by dismantling spitefully the me that I was for them" (p. 22). He chooses Marco Di Dio as the ``victim" for his next experiment.

Marco Di Dio is a local inventor in the town of Richieri. He aspired to great wealth and fame and received financial favors from both Moscarda and his father to keep his projects going. In addition, Moscarda's family allowed him to lodge in a rundown building on the edge of their property for many years, free of charge. In order to prove himself ``differently from one of the \textit{hundred thousand} in which [he] lived" (p. 74), Moscarda sets in motion a detailed plan to evict Di Dio and his wife from the building. The eviction turns out to be a bureaucratic nightmare but eventually Moscarda manages to toss Di Dio and his wife out into the street on a rainy winter day. A crowd of bystanders witness the eviction and sharply criticize Moscarda: ``More disgusting than his father!", ``In the rain, if you please! He wouldn't even wait till tomorrow!", ``Usurer! Usurer!". But then, to everyone's astonishment, Moscarda's lawyer announces that the evicted tenant will be donated one of Moscarda's more comfortable properties. With this unexplainable gesture, Moscarda earns the label of ``madman" but successfully executes his experiment, proving to others that he could be ``someone different from the man [he] was believed to be" (p. 95). This experiment is discussed at greater length and in terms of identity construction theory below, in the section titled ``The Marco Di Dio experiment".  

After the eviction of Marco Di Dio, Moscarda's closest family and acquaintances-Dida, Firbo and Quantorzo-fear that Moscarda will commit new follies. Returning home from a walk with his dog, Moscarda finds his wife Dida conversing with the bank manager Quantorzo in the living room. Fresh from a meditation about identity and the ``sight of the others" (p. 104), Moscarda realizes that there are not simply three people in the room (himself, Dida and Quantorzo). Rather, he recognizes eight different characters (p. 108): 

\begin{enumerate}[itemsep=-5pt]
\item Dida, as she was for herself;
\item Dida, as she was for me;
\item Dida, as she was for Quantorzo;
\item Quantorzo, as he was for himself;
\item Quantorzo, as he was for Dida;
\item Quantorzo, as he was for me;
\item Dida's dear Geng\`{e};
\item Quantorzo's dear Vitangelo.
\end{enumerate}

Moscarda does not include the missing ninth character-himself as he was for himself- having realized during his meditation that he was \textit{no one} for himself. In the eccentric discussion that follows (between these eight characters), Dida and Quantorzo try to persuade Moscarda to stay calm and respect the managerial decisions of Firbo and Quantorzo. But it is too late, as Moscarda, who had never involved himself with bank affairs in the past, has already decided to force the bank into liquidation-a gesture aimed at extinguishing the image that others had constructed of him: ``no longer usurer (that bank was done with!) and no longer Geng\`{e} (that marionette was done with, too!)" (p. 119). This discussion between Moscarda, Dida, and Quantorzo is revisited below in the section ``Conversing with Dida and Quantorzo".

The last two chapters of the novel (Books Seven and Eight) represent a narrative and stylistic departure from the rest of the story. Moscarda is invited to visit Anna Rosa, a friend of his wife that he hardly knows. While Moscarda is conversing with her, she inadvertently drops her handbag, setting off a revolver inside and wounding her foot. During her recovery, Moscarda visits Anna Rosa regularly and becomes very attached to her. He comes to realize that she too is uncertain of her ``authentic" identity and he introduces her to his tragedy and to his meditations on identity. Anna Rosa grows fascinated with Moscarda and his ideas to that point that, in a state of mental confusion, shoots him from her bed using the same revolver that had caused her injury (p.152): 

\begin{quote}
From that bed, a moment later, I rolled down blind, mortally wounded in the chest by the little revolver she kept under her pillow. The reasons she gave later in her defense must be true: namely, that she was driven to kill me by the instinctive, sudden horror of the act into which she felt she was drawn by the strange fascination of everything I had said to her during those days.
\end{quote}

Moscarda is taken to a hospital and upon recovering decides to ``set an exemplary and solemn example of repentance and self-sacrifice" (p.158) by donating all his possessions and joining a home for the destitute. When called upon to testify against Anna Rosa, Moscarda successfully absolves her of the crime. Once in the home, Moscarda is a transformed man. He has lost interest in public perceptions of his name and persona. He lives life moment by moment, leaving memories and names behind (p.160): 

\begin{quote}
And the air is new. And everything, instant by instant, is as it is, preparing to appear. [...] This is the only way I can live now. To be reborn moment by moment. [...] I die at every instant, and I am reborn, new and without memories: live and whole, no longer inside myself, but in every thing outside.
\end{quote}

These final events of the novel are revisited in the section titled ``Epilogue".

\section{The mirror experiment}

We begin our revisitation of Uno, Nessuno, Centomila with Moscarda's first identity experiment in which he attempts to watch himself ``live" in front of a mirror. This experiment is driven by Moscarda's desire to experience the image of himself from the gazing perspective of an outsider. There is a long tradition of linking identity troubles to mirrors and other reflective surfaces. Over a century ago, Cooley created the concept of the looking-glass self. He proposed that an individual constructs an understanding of her ``self" via others' perceptions in society, i.e., by imagining how she is perceived by others and developing an emotional response based on their judgments (Cooley, 1902). Mead (1934) further extended Cooley's theory, positing that the construction of the self happens entirely through social interaction. Reflexivity - experiencing oneself (as an object) from the standpoint of others - is still a core concept in Mead's theories, but the focus shifts to the interaction with the social world. Under the same rubric, and particularly relevant to the present discussion, is the dramaturgic work of Goffman (1959). In \textit{The Presentation of Self in Everyday Life}, Goffman posits the self as a malleable and elastic entity that is entirely shaped by the expectations formed by its interaction with the social world, the audience, and the stage. Again, the process of anticipating, interpreting, and assimilating the perception of others - similar to Moscarda's attempts to experience the outsider gaze - is a key notion in Goffman's work and is echoed in the work of contemporary theorists writing about online identity (Boyd, 2008; Pearson, 2008; Robinson, 2007; Van Kokswijk, 2009). We attempt to reframe this notion in a contemporary context, by posing the question: how would Moscarda perform the mirror experiment were he alive today? 

At the time at which Pirandello's story unfolds - the early twentieth century - cinematography was in its infancy and a mirror was the most readily available and convenient tool to experience the outsider gaze.  A contemporary Moscarda could make use of much more sophisticated technological devices. For the present discussion, we will leave high-tech stratagems aside - such as closed-circuit television, and illusory out-of-body experiences (Ehrsson, 2007) - and focus on what the outsider gaze experiment would entail regarding the present-day permeation of online social networks in our lives. Because modern identity issues necessarily extend beyond physical, offline fori, we speculate that a present-day Moscarda would be as concerned about the public perception of his physical persona as that of his virtual one. In other words, a Moscarda of the day would be curious to explore how he is portrayed and perceived both offline, in the real world, and online, on social media. 

It has been noted that the most common mechanism for presenting one's identity online on social networking sites is the personal profile (Boyd and Herr, 2006). A personal profile, is a web page consisting of personal information, such as name, occupation, location, relationship status, a diary, photographs, videos and other sorts of personal media. Despite the variety of services and options offered by different social networking sites the basic function of a profile is to present one's identity. A personal profile is thus the component of one's online identity that best approximates one's physical, public appearance.

We can then imagine that a present-day Moscarda would be curious to explore the ways in which he appears to others online. How could Moscarda achieve this? The simplest way would be by scrutinizing his own personal profile, attempting to imagine how others would perceive it. In other words, Moscarda would perform a modern version of the mirror experiment by logging into his social networking sites of choice, displaying, and perusing his personal profile page. An experiment that consists of merely scrutinizing one's own profile might appear trivial at first, but the highly malleable, dynamic, and real-time nature of social networking sites makes this experiment worth exploring more in detail. 

User profiles consist of variegated types of information. First and foremost, they contain personal information entered by their owners, such as name, a profile picture, relationship status, a biography, and photos. For example, Moscarda might have created a profile with information about his occupation (``banker"), his relationship status (``married to Dida"), and a profile picture. Moreover, he may choose to use the popular microblogging feature included in many social networks to share real-time news about himself as a brief text update (e.g., ``Moscarda is going to the bank"). A second category of information generally present on social network profiles relates to activities in the user social timeline. This includes information about recently established friendships (e.g., ``Moscarda is now friend with ***"), relationships (``Moscarda is now in a relationship with ***"), and social events (e.g., ``Moscarda attended the event  ***"). Many of these updates are triggered by users' interactions, e.g., accepting a friend request triggers an automated update. A third category of information found on user profiles is public content posted by third parties-friends, acquaintances, strangers, applications-in the form of text comments, ``wall posts" (profile annotations), tagged images and similar media content. For example, Dida might post a photo of Moscarda on his profile, making it publicly available and viewable to anyone visiting his profile (``Moscarda was tagged in a photo").  

It is important to differentiate between these categories because they constitute three different levels of authorship and three different mechanisms by which aspects of one's identity are revealed. The first category of information discussed above is under the direct control of the users, i.e., the owners of online profiles. The user carefully selects certain pieces of information to exhibit to the online world in order to fashion a desired image. As such, these exhibits may be subject to constant manipulation or curatorial effort. However, this form of identity construction is not exclusive to the online realm. As Moscarda (Pirandello, 1990:41) puts it:

\begin{quote}
Do you believe you can know yourselves if you don't somehow construct yourselves? Or that I can know you if I don't construct you in my way? And can you know me if I don't construct you in my way? We can know only what we succeed in giving form to.
\end{quote}

Similar to the ways by which people construct their personae and play with their own physical appearance in the offline world-by acting differently, by the application of make-up, by alternating attire, and even by undergoing plastic surgery-online appearance is also highly malleable: every morsel of information shared on a personal profile can be edited, replaced, modified, or concealed by its owners. This constant digital embellishment of one's profile points to the role of the social network as a performance stage, or a ``space for performing the self" (Chan, 2000:271). Users of online social networks ``perform" and construct an online identity via a constantly updated stream of text (microblogging messages, biographical notes, photo comments), videos, and images. Photographic images, in particular, are the most direct and powerful means to stage one's online performance. In Pirandello's days, photographic cameras were neither portable nor personal commodities. A present-day Moscarda, however, would certainly experiment with photographic devices to come to an understanding of his own self and appearance. Reflecting on the importance of the photographic medium in everyday life, Susan Sontag notes:  ``We learn to see ourselves photographically. To regard oneself as attractive is, precisely, to judge that one would look good in a photograph" (Sontag, 1979:85). Although Sontag was writing about analog photography, her words reverberate even more nowadays, for cameras are cheap, portable, and embedded in practically any electronic device. Cameras have become instrumental to digital aestheticization. With regard to this, it is interesting to note that a common practice among users of social networks, especially teenagers, is to take self-portraits using portable digital cameras in front of a mirror (Walker, 2007). Thus, the instrument employed by Moscarda to experience the outsider gaze becomes a tool of online identity construction and presentation.

While this first category of information consists of content that is directly under the control and management of users of social networks (name, biographical sketch, profile photos, status updates), the other two categories introduced above pertain to the social activities of users and are generated by third parties. These include content that is generated automatically by social networking sites (such as information about new friendship, the results of an online quiz, etc.) and other users (such as a new wall post from a friend, a new tagged photo, etc.). Although privacy measures exist (to different extents) on most social networking sites, by and large, information of this kind ends up on a user profile and is thus visible to anyone visiting it. 

In the pre-digital era, Moscarda could have kept this kind of information confidential. Of course, he could have shared it with a restricted circle of family and friends of his choice, but the bulk of his social activity would have gone largely undocumented, or confined to informal discourse and gossip. The systematic documentation of Moscarda's social whereabouts and activities (e.g., ``Moscarda is now friend with ***") together with the traces left on his profile by third parties (e.g., a wall post from a friend reading: ``hello Moscarda, it was great to see you yesterday at ***!") represents a crucial departure from the traditional ways in which one's identity is presented to the world. It follows that a Moscarda that operates both offline and online needs to control many more aspects of his identity to preserve a coherent rendering of his persona, many of which are outside of his direct control. 

Some preliminary conclusions emerge from the revisitation of this episode of Uno, Nessuno, Centomila. The offline Moscarda of the 1900s was concerned with the public perception of his persona and performed a mirror experiment to experience and analyze the outsider gaze. A contemporary Moscarda would find himself in a more complex landscape, for he would have to concern himself with public perceptions of both his offline and his online personae. We have imagined a modern Moscarda with an online profile on a social networking site and discussed the various kinds of identity-related information that would concern him in different ways. There would be information that he personally posts to his profile and information about his social whereabouts, activities and interaction with other users, posted by third parties. In this context, a modern version of the mirror experiment would force a present-day Moscarda to actively engage with his profile, manipulating it and curating it to his own desire. The ability to continuously and dynamically rework his profile would furthermore amplify Moscarda's anxiety about the presentation of his persona. With the aim of presenting a coherent version of himself to the world, Moscarda would find himself engaging in a dynamic process of identity construction through interaction. Thus, a modern version of the mirror experiment is a dynamic, real-time venture set in motion by our incessant preoccupation with the appearance and integrity of our ever-changing online profiles\footnote{We are increasingly becoming concerned with and alert to the status of our online identity. For example, it is not uncommon or unacceptable to rush to a computer to remove an embarrassing photo in which we were just tagged. Along these lines, we have heard of cases in which former couples meet (in person) to negotiate their ``break up on Facebook", i.e., to tweak the privacy settings on their profiles so to silence or attenuate the public, online announcement of their break up.}.

\section{The Marco Di Dio experiment}

While Moscarda's first experiment is merely a self-observation from the perspective of an outsider, in his second experiment, Moscarda intends to deconstruct every connotation associated with his persona. He delves into the ways he is perceived by others and finds that he abhors the identities that live within him (the dear Geng\`{e}, the ``son", the usurer). The intention to break with these identities (``I warn you, my friend, that I am not my father", p.88) is translated into a carefully crafted experiment in which he first intensifies his role of usurer, by publicly evicting Marco di Dio and his wife from their house, and then smashes it to pieces by moving them into a much more comfortable home, to the astonishment of everyone. 

Moscarda's second experiment is a social exercise of identity construction and deconstruction. Yet, how do we go about constructing identity in the first place? Let us turn to Jean-Paul Sartre and his speculation that a French caf\`{e} waiter is 'playing a waiter', not merely 'being a waiter': ``But what is he playing? We need not watch long before we can explain it: he is playing at \textit{being} a waiter in a cafe." (Sartre, 1943). As beings birthed into pre-existing societal constellations, we are outfitted with ready-made scripts and roles which we can choose to adopt, perform and even improvise on. J. L. Austin (1962) refers to these pre-scripted identities as ``performatives." In its original sense, the rubric ``performative" was intended to apply to certain ``illocutionary" speech acts that were neither true nor false, but ``performative." For instance, the illocutionary act: ``I pronounce you husband and wife," performs what it says provided that certain felicity conditions are met, e.g., the declarant has legal or religious capacity to do so, the partners are willingly present, and they fulfill certain legal and social prerequisites. Judith Butler (1999) expanded upon Austin's work to show that the term ``performative" can be extended to cover a sizeable host of habitually uninterrogated social roles and affectations. Though Butler concentrates her insights on gender (certain conditions must be met for one to be able to enact a certain gender), the same logical apparatus could just as well hold true for the performative script associated with police officers, mascots, the Sartrian waiter, as well as Moscarda. For example, in Uno, Nessuno e Centomila, the town of Richieri perceives Moscarda as a ``usurer" for he fulfils all the felicity conditions: (1) he is the owner of a bank that charges exorbitant interest rates (2) he lives by no other means and (3) he loafs about town in utter disoccupation. With his second experiment, Moscarda first evicts Marco Di Dio, and then rewards him with a more comfortable house, leaving the people of Richieri flummoxed. Public astonishment is what he had hoped for by breaking away from the ``usurer performative", by refusing to perform its script. How would Moscarda go about enact a similar performative construction and deconstruction in today's networked reality?

It is easy to imagine that a modern Moscarda would attempt to dismantle the public perception of his persona by tweaking his online performatives in unorthodox ways. There are innumerable stratagems by which he could catalyze an identity crisis, proving to others that he does not in fact embody certain pre-scripted identities: he could post embarrassing photos of himself or his friends, publish unusual, rude or politically incorrect comments, drastically change his profile information, publicly reveal personal secrets, or remove some of his crucial contacts. The list is potentially endless and not limited to a single social network. Modern social networking sites cater to different communities with different needs and different performatives. For example, LinkedIn caters to professionals looking for work, Match.com and similar dating sites cater to singles looking for partners, the elitist network asmallworld, with its coteries of yachting aficionados, organizes exclusive get-togethers and connection building for its members. MySpace, once the largest social network, has recently reinvented itself as a music and media industry stage. A Moscarda with profiles on these sites could play with his different performatives catalyzing site-specific identity crises. The availability and popularity of these many fori for identity construction and manipulation inflate the possibilities for specialized performative destabilization with ramifications both in the online and in the offline realm\footnote{Destabilizations of online performatives have created episodes of real life distress and hysteria. Examples on Facebook include losing a job over eccentric remarks on a status update (\url{http://www.hrzone.co.uk/item/171580}, and \url{http://bit.ly/aJFL2n}) and a man stabbing his wife to death after she changed her relationship status to ``single" (\url{http://news.bbc.co.uk/2/hi/7676285.stm}).}. In fact, one would think that the online social realm allows for more games and experiments of identity performance than its offline counterpart, for its immediate and dynamically constructed nature.

But this is not always the case. The popular social networking site Facebook is interesting to analyze in this context. With its sheer population size (over half billion users at the time of writing) and an unprecedented monopoly over users' personal data, Facebook is a forum in which multiple communities and societal roles necessarily meet: these days, your parents, your children, your colleagues, and your friends are all on Facebook. In such a diverse and highly populated environment, constructing, tweaking, and curating one's online representation is of crucial importance. Facebook, however, allows its users a very limited range of identity maneuver. By encoding prescriptive or formulaic alternatives within its system (gender: male or female; religious views: Christian, Jewish etc.; Political views: liberal, conservative, etc.), by slotting its users in preset geographical or associational networks, by enforcing the authenticity of user profiles, and by cloning everyone within the same spectrum of light blues and unadorned walls, Facebook regulates and limits its users' possibilities of representation. By contrast, Myspace gives its users full control over the HTML of their profile pages and allows them to fully re-imagine their profiles by redesigning backgrounds, colors, images, adding music, video streaming, and often clunky and discordant graphics. When the two sites are juxtaposed, Facebook's drastic restrictions on identity construction appear rather salient\footnote{Some have rebelled against the apparatuses of control and regulation posited by social media sites. For example, a Fakester Manifesto was written by people who crafted fake identities on the Friendster social network, ``in defense of our right to exist in the form we choose or assume", negating the possibility of authentic or 'real' profiles, and trying to extricate themselves from the webs of symbolic violence spun out on these sites. We note, however, that by fashioning aspirational profiles, the Fakesters for the most part merely endorse pre-scripted performative identities which, for a lack of felicity conditions, they are unable to carry off in the real world (a sort of Second Life fantasia). Moreover, by embracing a sort of consumer choice identity-matrix, a struggle in which only the most ``interesting" identities survive, the Fakesters would appear to actually reinforce those social cages which, like Moscarda, they hope to escape.}. 

Thus, even though the networked medium potentially enables myriad experiments of performative destabilization, identity games are limited by the techno-political boundaries designated by the administrators and owners of these social networking sites. In this light, one of the most fascinating phenomena on Facebook are its hortatory ``Take a Quiz" applications which permit individuals to engage in carefully delimited role playing and identity experimentation, while mitigating the risk that such performances would entail in the real world (as Moscarda's tragic fate demonstrates). The quizzes are manifold in subject matter, running from \textit{Which Desperate Housewife are you?} to \textit{Which German Philosopher are you?} and \textit{If you were a famous serial killer, who would you be?} The results are always quirky, ludic and normative, and they legitimize identity experimentation at the same time that they quell Pirandellian fears by providing authoritative scripts. Also, their exclusions become less obvious over time. If for Butler, mimicry and masquerade form the essence of identity, then Facebook offers a padded playpen in which to explore the polyglot nature of the self, while at the same time homogenizing its adherents by excluding the radical and the troubling\footnote{This takes on a sinister air when one delves a little into the political and social stances of Facebook's original backers. Tom Hodkinson in his searing article in The Guardian, ``With Friends Like These'' (\url{http://gu.com/p/xx49p}), singles out Peter Thiel, Zuckerberg's original underwriter, and one of Facebook's three board members. He notes: ``The real face behind Facebook is the 40-year-old Silicon Valley venture capitalist and futurist philosopher Peter Thiel, [...] more than just a clever and avaricious capitalist, he is a futurist philosopher and neocon activist. A philosophy graduate from Stanford, in 1998 he co-wrote a book called The Diversity Myth, which is a detailed attack on liberalism and the multiculturalist ideology that dominated Stanford. He claimed that the "multiculture" led to a lessening of individual freedoms." Hodkinson warns that Thiel's philosophy and business agendas seek to scour away social and cultural difference and give rise to a homogenized and fungible global populace.}. No matter how wildly Moscarda responded to the quiz questions, he would unavoidably find that his personality matched up with some eccentric but palatable German philosopher, vivacious Desperate Housewife or laughable serial killer. 

More worringly perhaps, as Facebook gradually monopolizes the social networking world market, its identity reductionism rubric slowly fades from view. As Butler (1999:5) writes, ``juridical subjects are invariably produced through certain exclusionary practices that do not ''show" once the juridical structure of politics has been established." In an Orwellian fashion, taxonomies that are no longer available become unthinkable, and users learn to desire only what conditions make possible for them. Without the presence of alternative social networks to differentiate from Facebook, it becomes difficult to even imagine those possibilities that have been excluded.

\section{Conversing with Dida and Quantorzo}
After the public debacle with Marco di Dio, Moscarda goes through another phase of self-unravelment, through which he comes to realize that he inhabits countless identities - as many as are the people that know him or know of him. The climax of this realization comes with the eccentric discussion between Dida and Quantorzo in which Moscarda recognizes not only the three people conversing at the table (Dida, Quantorzo, and himself), but an additional five characters predicated upon the reciprocal perception of their identities (Dida as she was for Quantorzo, Dida as she was for him, etc.)

Sociologist Pierre Bourdieu's notion of ``habitus", or embodied disposition (Bourdieu, 1977), is helpful to analyze the confrontation between Moscarda, Dida, and Quantorzo and its resulting discomforts. Bourdieu posits that individuals enter and subsequently exist in different fields of activity, each of which bears its own complex structure of social relations. Depending on their position in a given social field (dominant or dominated, outsider or insider, etc.) they develop a certain habitus typical of their position: not merely an assemblage of dispositions, opinions, taste, and aesthetic sensibility, but also certain mannerisms, gestures, and tones of voice. Since each social field places requirements on its participants for membership within the field (certain dress, posture, and speech), actors vying for membership tend to construct their habitus according the given social organization of the field. In some respects, Bourdieu's postulates are consonant with Erving Goffman's ideas on symbolic interactionalism: personal identity comes into being via social interaction, and according to the different roles we play in different contexts. 

In a multi-field world, actors routinely adopt and shed habituses depending on their different standings in each field. Trouble and discomfort arise when two different fields overlap synchronically, and a particular actor is unsure what habitus to adopt. This clash of habituses is emblematically portrayed in Uno, Nessuno, Centomila in Moscarda's confrontation with Dida and Quantorzo. In the midst of his identity crisis, Moscarda finds himself anxiously caught in a room with his wife, with whom he wear one habitus (``Geng\`{e}", the submissive and almost nebbish husband), and his friend and employee Quantorzo, with whom he wears a different one (that of Vitangelo, authoritative bank owner and son of the renowned town usurer). 

The opportunities for a clashing of habituses - which in Moscarda's offline world amount to a discrete if discombobulating confrontation - are manifold in the online realm. Social media sites offer open, public, and constantly accessible forums for confrontations of this kind - imagine a respected colleague, a childhood friend, and a close relative posting wildly incongruent comments in response to an update on your Facebook wall. These days, the discussion between Moscarda, Dida, and Quantorzo could unfold on a public online forum, in a private message exchange, or as a distributed game of photo tagging and item commenting. Moreover, social network spaces act as powerful vehicles for the display and transmission of social preconceptions and assumptions similar to the ways that one presents one's habitus offline. Would Moscarda present the same profile picture of himself on LinkedIn, Myspace, and Match.com? Probably not.\footnote{It is easy to argue with this view. The notion of habitus is not merely about socialization. A critical feature of habitus is also its embodiment. Habituses are embodied in one's bodily gestures, mannerisms, and verbal attitude. Thus, one could argue that embodied verbal and visual clues are probably not transmitted online as well as they are offline, so that the emergence of online habituses is somehow limited by the narrow range of bodily practices attainable on the networked medium. However, recent research shows that when online venues preserve the dynamics of interactional cuing, embodied offline cuing systems are transmitted and redefined online: ``In creating online selves, users do not seek to transcend the most fundamental aspects of their offline selves. Rather, users bring into being bodies, personas, and personalities framed according to the same categories that exist in the offline world." (Robinson, 2007:94)} By corollary, his entire field of activity - the status of his habitus within a particular online universe - would be different in each of these sites. Not only do social networking sites offer a forum for habitus confrontation but they also provide the structural and social conditions for the emergence of novel ad hoc habituses. 

With these notions in mind, we can then imagine that a present-time conversation between Moscarda, Dida, and Quantorzo is animated by many more characters than the eight posited by Pirandello in a purely offline world. Let us suppose, for example, that Dida has an online profile on the social networking site Facebook and that both Moscarda and Quantorzo are familiar and keep up to date with it. If this were the case, by entering the conversation, Dida would be bringing to the table not only her two offline habituses (Dida, as she is with/for Moscarda, and Dida, as she is with/for Quantorzo), but also an additional character predicated upon her online ``field of activity" (Dida as she is on Facebook). With the presence and availability of several social networking sites, each giving the opportunity to develop ad hoc habituses, it is easy to realize that the number of characters at play is potentially very large (Dida as she is on Myspace, Dida as she is on asmallworld, etc.). As noted by Van Kokswijk (2008), this identity proliferation does not necessarily undermine the integrity of one's ``real" identity. Rather, he contends that by having different profiles and wearing different habituses, Dida (or anyone) does not decentralize or diminish her identity; rather, she multiplies it infinitely.
 
To add complexity to an already complex scenario, we should not forget that social networking sites allow users to customize their privacy settings so that their online profiles appear differently to specific people. Users are able to customize every aspect of their personal profile, e.g. restricting certain portions of their profiles to certain contacts and making other parts publicly available. This means that, at any time, a person's profile might appear different to two different profile visitors that have been granted distinct access privileges. This has immediate repercussions for the clash of habituses discussed above. Dida's field of activity on Facebook will not result anymore in a unitary habitus (Dida as she is on Facebook), but of many different ones (Dida as she is on Facebook to Moscarda, and Dida as she is on Facebook to Quantorzo).

In this rich ecology of online personae, navigating through a plethora of privacy settings and access privileges is not a trivial task. Interestingly, to aid management of such complex and distributed identity configurations, Facebook allows you to ``see how a friend sees your profile", i.e., to display your profile exactly the way it is seen by a given friend based on the selected privacy restrictions (e.g., Moscarda could  ``see how Dida sees his profile"). Playing with this functionality corresponds, in many respects, to performing a customized version of the mirror experiment discussed earlier: experiencing not a generic outsider gaze, but that of a specific outsider. 

\section{Epilogue}
The final chapters of Uno, Nessuno, Centomila narrate the complete breakdown and unraveling of Moscarda's personality. Via his carefully crafted experiments, Moscarda manages to dismantle the multiple identities he was associated with: ``no longer usurer (that bank was done with!) and no longer Geng\`{e} (that marionette was done with, too!)" (p. 119). But the search for his veritable self does not turn out to be what he expected. The deconstruction of the mosaic of personae that lived within him ultimately leaves him with a void: ``But what other did I have inside me, except this torment that revealed me as no one and one hundred thousand?" (p. 120). 

The chain of unruly and tragic events that follows - with Anna Rosa shooting her foot and then nearly murdering Moscarda - is a metaphor for the latent but always potential violence that swims beneath the glassy surface of social relations, i.e. the violence that meets those who seek to disrupt social conventions. The story culminates with Moscarda surrendering to his own tragedy. He chooses to entirely extricate himself from his social fabric and to live contentedly by himself in a house for the destitute. There is no real catharsis to this tragedy. Moscarda does not find his true self and abandons any attempt to do so: ``I no longer look at myself in the mirror, and it never even occurs to me to want to know what has happened to my face and to my whole appearance" (p. 159). By completely becoming an outsider, he no longer has to pursue being an outsider to himself in order to probe his identity. The analogy with today's world is obvious: the disintegration of Moscarda and his renunciation of a social life is analogous to permanently logging off a social network. 

However, logging off, quitting, or even temporarily abandoning social networks is not easy. Sharing different aspects of one's life on social networks has become routine for many. In many respects, social networks have become modern platforms for autobiographical expression. When he inaugurated the modern autobiography with his work \textit{Confessions}, Jean-Jacques Rousseau (1782) claimed that to communicate one's true identity, one must ``tell all." ``Such as I was," he writes, ``I have declared myself; sometimes vile and despicable, at others, virtuous, generous and sublime; even as thou hast read my inmost soul." Rousseau's stance can be profitably contrasted with the circumspect revelations of identity one encounters in online social networks. As discussed above, social network users endeavor only to show ``part": to slice and repackage their identities so as to manifest only their most attractive aspects. This online cropping often reflects similar stratagems in the analog world: ``putting on one's best face at work," or attentively trying to make a ``good first impression." 

Yet, these efforts are counteracted by the thoroughly dynamic, immediate and interactive nature of social networking sites that tacitly or often explicitly coerce their users to constantly act upon their social circles: ``Unlike everyday embodiment, there is no digital corporeality without articulation. One cannot simply 'be' online; one must make one's presence visible through explicit and structured actions." (Boyd, 2008:149). In this vein, most Facebook users are incessantly prompted to contact friends who they have not been in touch with lately (``Write on ***'s wall! Send her a message!"), and to browse through endless lists of suggested friends (``Here are some people you may know"). By incessantly reminding its users to send messages or friend requests, to write on walls, to ``like" or ``not like," to ``poke," to ``give gifts," to choose role models via \textit{Take a Quiz} and to engage in a whole other range of formulaic activities, Facebook spurs users to act upon their profiles and their online social environment. In many respects, these are the ways by which Facebook recultures and educates its users via its very own forms of ``symbolic violence", to use Bourdieu's terminology. Thus, in the long run, despite any user's most resourceful exertions, the social network machine itself soon begins to ``tell all." The Facebook panopticon is such, with its ceaseless feeds, updates, tags, pokes, gifts, city locators, and quiz results, that the flux and hydra-headed polyphony of the self leaches onto the screen with a speed and nimbleness that overwhelms any possible human counter-stratagems. 

How could a modern Moscarda extricate himself from such a tangled and resilient social fabric? Even committing a ``digital hara-kiri" - by deleting his account on every social networks he belongs to - would not guarantee his escape from the online social world: profiles on some social networking sites (including Facebook) cannot be erased. At the most, they can be deactivated. As such, our imagined digital Moscarda, by signing up on Facebook, unwittingly subscribed to a multimedia biography, written by hundreds of thousands of authors, that will outlast himself.  

\section{No conclusion\marginnote{We pay homage to the work of Pirandello by titling the concluding section of this article ``No conclusion", as the last section of the book. Accordingly, we offer no definite and fixed conclusion to our article. Rather, we would like to leave this forum open for discussion as we urge other scholars to embrace Pirandello's ideas in their work on online identity and performance. }}

\newpage

\section{References}
\begin{enumerate}[itemsep=-2pt]
\small
\item Austin, J. L. (1962) \textit{How to do Things with Words: The William James Lectures delivered at Harvard University in 1955.} Oxford: Clarendon. 
\item Bourdieu, P (1977) \textit{Outline of a Theory of Practice}, Cambridge: Cambridge University Press.
\item Boyd, D. and Heer, J. (2006) 'Profiles as Conversation: Networked Identity Performance on Friendster' paper presented at the Hawai'i International Conference on System Sciences (HICSS-39), Kauai, HI.
\item Boyd, D. (2008) 'None of this is Real', in J. Karaganis (ed.) \textit{Structures of Participation in Digital Culture}, pp. 132-157. New York: Social Science Research Council. 
\item Butler, J. (1999) \textit{Gender trouble: feminism and the subversion of identity}. Routledge.
\item Chan, S. (2000) 'Wired\_Selves: From artifact to performance' \textit{CyberPsychology and Behavior}, 3(2):271-285.
\item Cooley, C.H. (1902) \textit{Human Nature and the Social Order}. New Brunswick, NJ:Transaction.
\item Ehrsson, H. (2007) 'The Experimental Induction of Out-of-Body Experiences', \textit{Science}, 317(5841):1048.
\item Goffman, E. (1959) \textit{The Presentation of Self in Everyday Life}. New York: Anchor Books.
\item Mead, G.H. (1934) \textit{Mind, Self, and Society}. Chicago, IL: University of Chicago Press.
\item Pearson, E. (2009) 'All the World Wide Web's a stage: The performance of identity in online social networks', \textit{First Monday}, 14(3).
\item Pirandello, L. (1990) \textit{One, No one and One Hundred Thousand} (Translated by William Weaver). Boston, MA: Eridanos Press, Inc.
\item Robinson, L. (2007) 'The Cyberself: the Self-ing Project goes Online, Symbolic Interaction in the Digital Age' \textit{New Media and Society}. 9(1):93-110. 
\item Rousseau, J. (1782) \textit{Confessions}.
\item Sartre, J. (1943) \textit{Being and Nothingness. A Phenomenological Essay On Ontology}. Gallimard.
\item Sontag, S. (1979) \textit{On photography}, London: Penguin Books.
\item Van Kokswijk, J. (2008) 'Granting personality to a virtual identity', \textit{International Journal of Humanities and Social Sciences} 2(4).
\item Walker, J. (2007), 'Mirrors and Shadows: The Digital Aestheticisation of Oneself', paper presented at the Digital Arts and Culture Conference, Perth, Australia.
\end{enumerate}
 
\end{document}